\documentclass[aps,prd,twocolumn
,showpacs,preprintnumbers,superscriptaddress,floatfix]{revtex4}

\usepackage{graphicx}  
\usepackage{dcolumn}   
\usepackage{bm}        
\usepackage{amssymb}   
\usepackage{setspace}
\usepackage{amsmath, amssymb, setspace}
\usepackage{array}
\usepackage{booktabs}
\usepackage{caption}
\usepackage{mathrsfs}
\usepackage{indentfirst}
\usepackage{slashed}
\usepackage{float}
\usepackage{lmodern}
\usepackage{hyperref}
\usepackage{xcolor}
\usepackage{multirow}
\usepackage{epstopdf}
\newcommand\Tstrut{\rule{0pt}{2.6ex}}         
\newcommand\Bstrut{\rule[-0.9ex]{0pt}{0pt}}   

\begin{document}

\title{On particle production in jets with quark-like and gluon-like fragmentation}

\author{Sona Pochybova}
\email[]{sona.pochybova@cern.ch}
\affiliation{Wigner RCP, HAS \\Konkoly-Th\'ege Mikl\'os \'ut 29-33, 1121 Budapest, Hungary}
\date{\today}

\begin{abstract}
We study the $p/\pi$ ratio in jets produced in simulated proton-proton collisions at $\sqrt{s_{NN}}=7~\mathrm{TeV}$ using Pythia $6.4$ Monte-Carlo generator. We compare the $p/\pi$ ratio in the selected quark-like and gluon-like jets to a reference samples of quark- and gluon-jets tagged at Monte-Carlo level. We observe that the contamination in the selected jets significantly influences the observed ratios. This suggests, that the origin of the jet fixes the value of the $p/\pi$ ratio within the model that we use.
\end{abstract}

 
\maketitle

\section{Introduction}

At LHC, partons inside hadrons colliding at high energies may experience hard scatterings. This kind of interaction produces correlated showers of particles, that are experimentally identified as jets. We identify two type of partons as origins of such showers: quarks and gluons. The basic $2\to2$ hard scattering processes are summarized in Tab.~\ref{tab:tab1}\cite{Sjostrand:2006za}:

 \begin{table}[h!b]
 \label{tab:tab1}
 \begin{centering}
 \begin{tabular}{ll|ll}
 \hline
 \textbf{Process ID}\Tstrut\Bstrut & Process & \textbf{Process ID}\Tstrut\Bstrut & Process
 \Tstrut\Bstrut \\
 \hline
 \textbf{11}\Tstrut\Bstrut & $q_{i}q_{j}\-\to q_{i}q_{j}$ &\textbf{28}\Tstrut\Bstrut & $q_{i}g\-\to q_{i}g$ \Tstrut\Bstrut \\
 \hline
 \textbf{12}\Tstrut\Bstrut & $q_{i}\bar{q_{i}}\-\to q_{k}\bar{q_{k}}$ &\textbf{53}\Tstrut\Bstrut & $gg\-\to q_{k}\bar{q_{k}}$ \Tstrut\Bstrut \\
 \hline
 \textbf{13}\Tstrut\Bstrut & $q_{i}\bar{q_{i}}\-\to gg$ & \textbf{68}\Tstrut\Bstrut & $gg\-\to gg$ \Tstrut\Bstrut \\
 \hline
 \end{tabular}
 \caption{Basic QCD jets processes with their IDs in PYTHIA}
 \end{centering}
 \end{table}

The properties of the jet we observe is determined by how the original parton fragments along its way. There are differences between quarks and gluons.  These differences are theoretically embraced in the QCD Casimir factors (also known as color factors), which are proportional to the probability that a parton radiates a soft gluon. Gluon's color factor ($\mathrm{C_{A}}$) is more than twice larger than that of a quark ($\mathrm{C_{F}}$) \cite{Ellis:1991qj}:
\begin{equation}
\frac{C_{A}}{C_{F}}\ =\ \frac{3}{4/3}\ =\ 2.25, 
\end{equation}
which means that gluons are expected to form higher multiplicity jets with softer fragments distributed in a larger jet-cone.

The differences between quark- and gluon-jets were tested extensively at LEP in $e^{+}e^{-}$ collisions\cite{Abreu:1995hp} and later at Tevatron in $p\bar{p}$ collisions \cite{Affolder:2001jx}. In both experiments the above expectations have been fulfilled. Furthermore, at LEP  the $C_{A},\ C_{F}$ factors have been measured to be $C_{A} =2.89+-0.01(stat.)+- 0.21(syst.)\ \mathrm{and}\ C_{F}=1.30+-0.01(stat.)+-0.09(syst.)$. These are consistent with the QCD predictions \cite{Kluth:2003yz}. 

At LEP it was also observed that gluon-jets on average produce more protons than quark-jets \cite{Abreu:2000nw}. Of course, this finding has been included into the parameters of the fragmentation functions we use in Monte-Carlo (MC) simulations \cite{Albino:2005me}. On the other hand, this effect was not explained theoretically which opens possibilities for an investigation of this subject. In our previous work\cite{Pochybova:2010gf} we studied particle production in gluon- and quark-enriched events. In this work, we designed an exercise, to see, what is the crucial factor in a MC model, that determines the $p/\pi$ ratio in a jet. Is it the nature of the leading parton, or is it the way that a jet object fills the phase space with particles? 

First step to be taken to answering this question, is to obtain samples of quark- and gluon-jets. There are efforts put into identifying jets as quarks and gluons at LHC. A thorough theoretical approach was taken in this matter in \cite{Gallicchio:2011xc}. The cited work aims at obtaining high-purity samples. We decided to take a rather different approach. In order to understand where the excess of protons in gluon-jets comes from and to answer the above question, we allow for contamination on purpose and observe how the final ratios change with respect to reference samples of MC quark- and gluon-jets.

To summarize, the aim of our study is to see, how the observed differences between quark- and gluon-jets determine the final identified particle spectra within a widely used MC model. Further we want to motivate a study on data, that divides the jets into two samples with either quark-like or gluon-like fragmentation. On comparison with reference samples of quark- and gluon-jets, we can conclude whether the resulting $p/\pi$ ratio is determined by the way the jet fragments or its origin. Our aim is not to compare different MC models. Instead we want to introduce a way to look at jets, that can shed more light on why gluon-jets produce more protons relative to quark-jets. If it is the origin that matters, the contamination will significantly affect the final particle spectra. If it is the fragmentation that is important, we should not observe difference within a certain fragmentation class, no matter the contamination. 

\section{The separation method}

In this section we introduce a method to separate the jets into two fragmentation classes. We do this by looking at the energy distribution inside a jet and the respective number of tracks. To obtain the cuts we use 3--jet and $\gamma$--jet events. We chose to work with these, since they provide samples of quark- and gluon-jets as they are their experimental sources. 

\subsection{Data sample and event selection}

To separate our sample we use the sets of 3--jet and $\gamma$--jet events obtained from simulated proton-proton(pp) collisions at $\sqrt{s_{NN}} = 7~\mathrm{TeV}$. This way we calibrate the cuts used to distinguish between quark-like and gluon-like jets. In order to obtain these sets, we generated $100~\textrm{milion}$ events containing QCD hard processes and $60~\textrm{milion}$  events containing direct-photon production. For our study we use $\mathrm{Pythia}~6.4$ MC generator\cite{Sjostrand:2006za}, Perugia 0 tune\cite{Skands:2010ak}. To reconstruct jets we use the $\mathrm{anti-k_{T}}$ jet-finding algorithm \cite{Cacciari:2008gp} with the following parameters: $R=0.4$, $p_{T}^{jet}>5~\mathrm{GeV/c}$, $|\eta_{jet}|<0.5$ and $|\eta_{particle}|<0.8$. We have used the same set of kinematic cuts that are used by the ALICE collaboration at LHC. We have made such a decision, since we are interested to study identified particles inside jets and ALICE provides the necessary particle-identification capabilities to do so. 

Further, we divided these samples into a sample containing 2 leading jets from 2-- and 3--jet events, a sample containing the least energetic jet from 3--jet events and finally a sample of jets from $\gamma$--jet pairs. In order to obtain these samples, we imposed specific event-selection criteria.

\paragraph{Selection of 2--jet events} What we are looking for in a 2--jet event is a pair of jets, which are well balanced and produced in plane. In order to do so, we selected events with two reconstructed jets and aplanarity\cite{Sjostrand:2006za} less than $0.01$.

\paragraph{Selection of 3--jet events} What we are looking for in a 3--jet event is a triplet of jets, which are produced out of event-plane and take a topology, which is close to a 'Mercedes-Star like' one. In order to do so, we require aplanarity to be greater than $0.05$ in events, where 3 jets were reconstructed. We consider the least energetic jet to be a gluon.

\paragraph{Selection of $\gamma$--jet events} For a $\gamma$--jet event we are looking for an event with a pair of direct photon and a jet with the following properties: $\textrm{aplanarity}\-<\-0.006$, $p_{T}^{imbalance}\-<\-0.1$\footnote{$p_{T}^{imbalance}=(p_{T}^{\gamma}-p_{T}^{jet})/p_{T}^{\gamma}$} and $\Delta \varphi_{\gamma-jet} \in (2.8;3.4)$.

\paragraph{}By these selections we acquired a sample of mixed quark-- and gluon--jets (leading jets from the 2-- and 3--jet events, further QG(MC)), that will serve as a pool for our selection, sample of gluon-jets (least energetic jet from the 3--jet event sample, further G(3J)) and finally, a sample of quark-jets from the $\gamma$--jet events (the single jet from $\gamma$--jet pair, further Q($\gamma$J)). A table with abbreviations of different jet-samples to be used further in the text can be found in Tab.~\ref{tab:tab2} on Page~\pageref{tab:tab2}. 

 \begin{table*}[!t]
 \centering
 \caption{Table explaining the naming convention in the text and legends of figures}
 \label{tab:tab2}
 \begin{tabular*}{\textwidth}{@{\extracolsep{\fill}}ll@{}}
 \hline
 \textbf{QG(MC)}\Tstrut\Bstrut & 2 leading jets selected from 2-- and 3--jet events from the generated jet sample; no process is selected\Tstrut\Bstrut \\
 \hline
 \textbf{G(3J)}\Tstrut\Bstrut & gluon--jets selected from 3--jet events\Tstrut\Bstrut \\
 \hline
 \textbf{Q($\gamma$J)}\Tstrut\Bstrut & quark--jets selected from $\gamma$--jet events\Tstrut\Bstrut \\
 \hline
 \textbf{Q(MC)}\Tstrut\Bstrut & 2--jet events selected from events in which only quarks are present in final state of $2\to2$ scattering\Tstrut\Bstrut \\
 \hline
 \textbf{G(MC)}\Tstrut\Bstrut & 2--jet events selected from events in which only gluons are present in final state of $2\to2$ scattering\Tstrut\Bstrut \\
 \hline
 \textbf{Q(sel)}\Tstrut\Bstrut & jets passing the cuts to select quark--like jets\Tstrut\Bstrut \\
 \hline
 \textbf{G(sel)}\Tstrut\Bstrut & jets passing the cuts to select gluon--like jets\Tstrut\Bstrut \\
 \hline
 \end{tabular*}
 \end{table*}

\subsection{Selection of quark- and gluon-like jets}

In this subsection we introduce a set of cuts that will be used on the QG(MC) to select quark-like and gluon-like jets. These set of cuts are designed to select jets with similar fragmentation to that of G(3J) and Q($\gamma$J).

\begin{figure*}
\centering
\includegraphics[width=\textwidth]{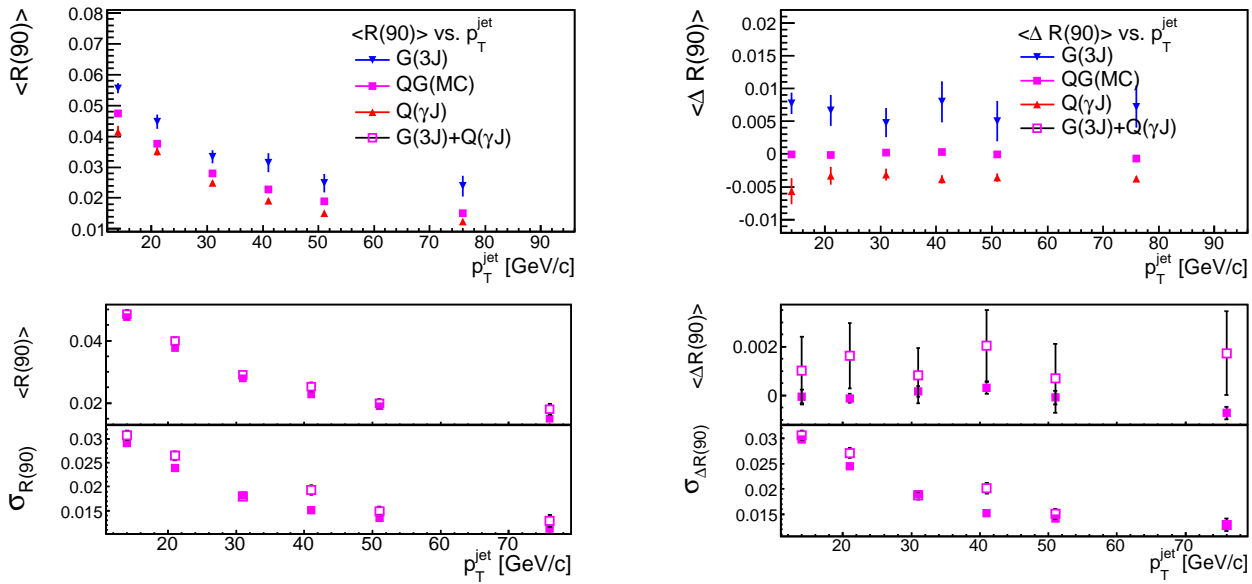}
\caption{Left: $R(90)$ as a function of jet's momentum for different jet-selections(top). In the two bottom panels the sum of G(3J) and Q($\gamma$J) is compared to QG(MC) in terms of average values and widths. Right: $\Delta R(90)$ as a function of jet's momentum for different jet selections(top). In the two bottom panels the sum of G(3J) and Q($\gamma$J) is compared to QG(MC) in terms of average values and widths.}
\label{fig:fig2}
\end{figure*}

\paragraph{$R(90)$ and $\Delta R(90)$ variables}
At first we looked at how the energy is distributed in different jets based on which sample they belong to. This we did using a jet-shape-like variable, which we call $R(90)$. It is the sub-cone size, which contains $90\%$ of jet's momentum. The distribution of $\langle R(90)\rangle\-vs.\-p_{T}^{jet}$ is shown in Fig.~\ref{fig:fig2}. We see that the value for G(3J) is higher than for Q($\gamma$J), as expected from the quark and gluon difference, and further the value for QG(MC) lies between these two, suggesting, that it is a mixture of quarks and gluons. However, to be able to use the G(3J) and Q($\gamma$J) samples to introduce a selection cut, we need to show that indeed, on combination of the G(3J) and Q($\gamma$J), we obtain the same distributions of $R(90)$ as with QG(MC). This comparison is shown in Fig.~\ref{fig:fig2} in the bottom of left panel, where the average values and widths of $R(90)$ are compared for QG(MC) and Q($\gamma$J)+G(3J). On combination, the average values and widths are comparable.

We see, that the $R(90)$ depends on jet's momentum. To reduce this dependence we introduce $\Delta R(90)_{\{G(3J);Q(\gamma J)\}}=R(90)_{\{G(3J);Q(\gamma J)\}}-\langle R(90)\rangle _{QG(MC)}$ at a given $p_{T}^{jet}$ momentum bin. This way we obtain the distribution in Fig. \ref{fig:fig2}, right panel. We see that we have reduced the momentum dependence and can thus work in a wider range of momenta. We chose to work with the jets $p_{T}^{jet}\in(16;56)~\mathrm{GeV/c}$.

\begin{figure}[!h]
\begin{minipage}{\columnwidth}
\centering
\includegraphics[width=\columnwidth]{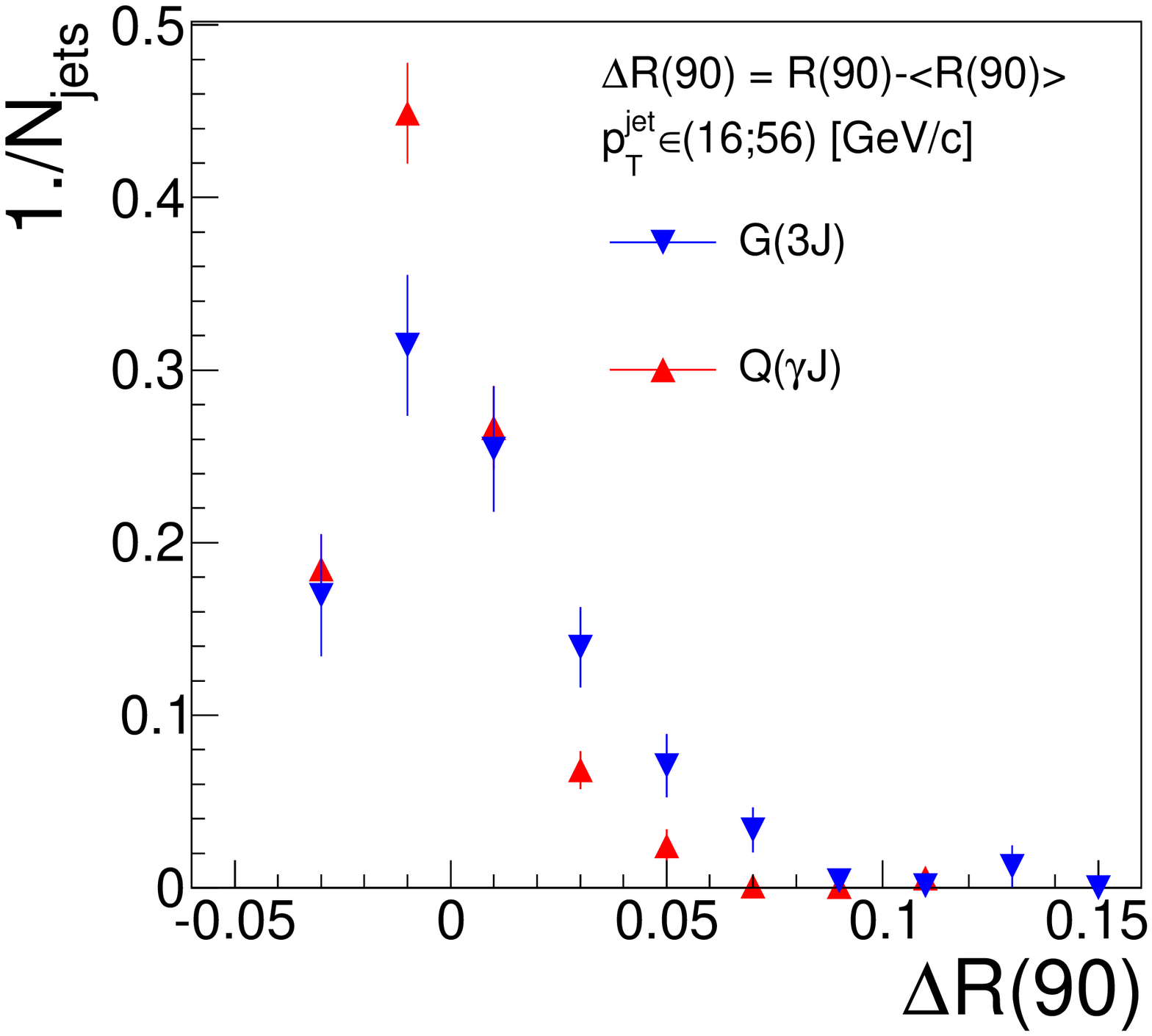}
\end{minipage}
\caption{$\Delta R(90)$ distribution for G(3J) and Q($\gamma$J).}
\label{fig:fig3a}
\end{figure}

The distribution of $\Delta R(90)$ for G(3J) and Q($\gamma$J) in given momentum range is shown in Fig.~\ref{fig:fig3a}. We can distinguish $\Delta R(90)$ intervals in which either the G(3J) or Q($\gamma$J) dominate, although they are overlapping. We select the following cuts for quark-like and gluon-like jets: $\mathrm{G(sel)}:\-\Delta R(90)\in (0.02, 0.04)$, $\mathrm{Q(sel)}:\-\Delta R(90)\in (0.02, 0.04)$.

\begin{figure}[!h]
\begin{minipage}{\columnwidth}
\centering
\includegraphics[width=\columnwidth]{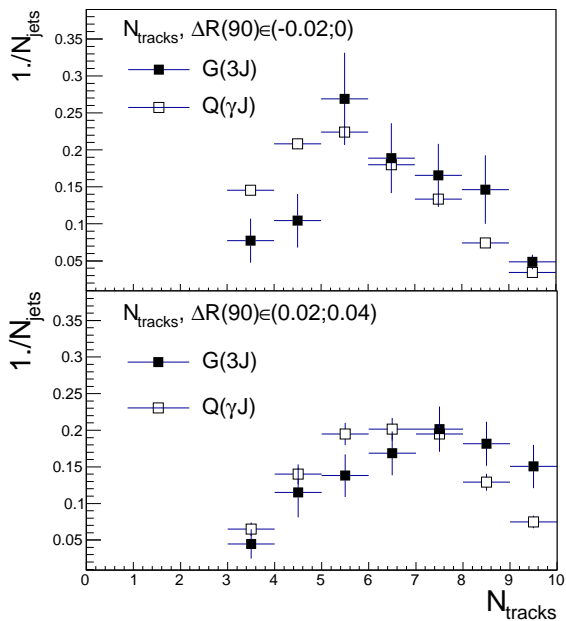}
\end{minipage}
\caption{Number of tracks distribution for G(3J) and Q($\gamma$J) events in selected $\Delta R(90)$ intervals.}
\label{fig:fig3b}
\end{figure}

\paragraph{Number of tracks} We want to obtain 2 samples of jets so, that each of them contains jets with similar, quark-- or gluon--like fragmentation. The first step is to select jets with similar energy distribution inside a jet--cone as described in previous paragraph. The second step is to look at the number of tracks inside each jet in the chosen $\Delta R(90)$ sub-intervals based on whether this jet comes from G(3J) or Q($\gamma$J). The distributions of the number of tracks inside jets are shown in Fig. \ref{fig:fig3b}. We see, that based on which sample the jet came from, the distribution of the number of tracks varies according to expectations, even in a narrow $\Delta R(90)$ bin. This means, that Q($\gamma$J) have smaller number of tracks than G(3J). To enhance the separation of the jets into two samples of different fragmentation, we apply an additional cut on the number of tracks in a jet: $\mathrm{G(sel)}:\-N_{tracks}=8$, $\mathrm{Q(sel)}:\-N_{tracks}=3$. The selection we have made, allows for contamination of the selected samples. As mentioned earlier, we do not want to get rid off this contamination, rather we want to see, how the contamination will influence particle spectra inside jets. 

\section{Comparison of selected quark-- and gluon--like jets}

In this section we compare the $p/\pi$ ratios of the selected jets to the 3--jet and $\gamma$--jet samples and MC quark--jets (Q(MC)) and MC gluon--jets (G(MC)).

\begin{figure*}[htb]
\centering
\includegraphics[width=\textwidth]{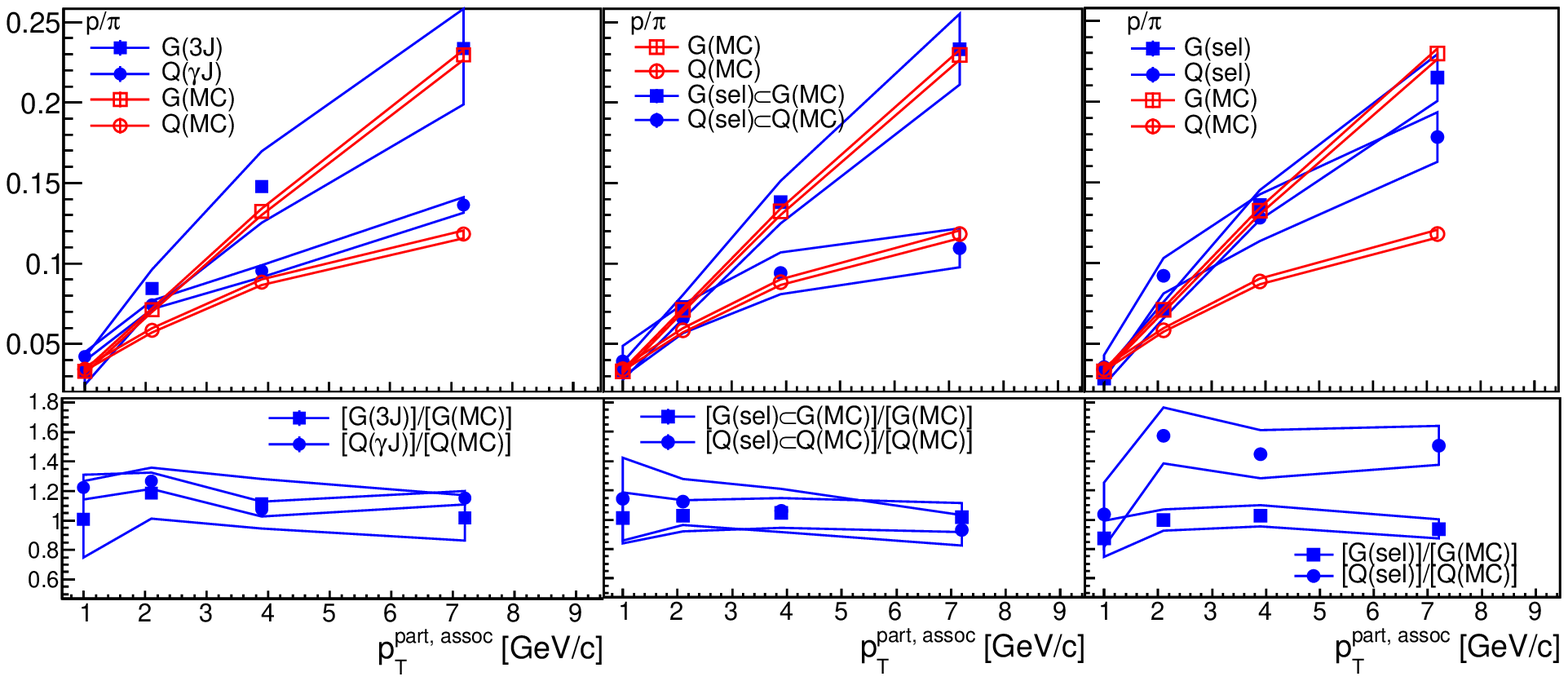}
\caption{$p/\pi$ ratio compared for different jet selections. Left: $p/\pi$ ratio for $\gamma$--jet, 3--jet and Q(MC) and G(MC). Middle: $p/\pi$ ratio for a subset of Q(MC) and G(MC), passing the criteria selecting quark-- and gluon--like jets, respectively, compared to Q(MC) and G(MC). Right: $p/\pi$ ratio for Q(sel) and G(sel) compared to the one for Q(MC) and G(MC).}
\label{fig:fig4}
\end{figure*}

First we compare the Q($\gamma$J) and G(3J) samples to Q(MC) and G(MC). The MC jet--samples were obtained by simulating events with hard scatterings  producing either quarks or gluons in the final state. These processes and their respective process IDs can be found in Tab.~\ref{tab:tab1} on page~\pageref{tab:tab1}. Subsequently, in these events we ran the jet--finding algorithm and selected 2--jet events as described in the previous section. The comparison is shown in Fig.~\ref{fig:fig4}, in the very left panel.

As can be seen, the $p/\pi$ ratios between the respective samples of quark-- and gluon--jets are comparable and our assumption to use G(3J) and Q($\gamma$J) for studying the influence of quarks and gluons on the $p/\pi$ ratios is justified within the used MC model.

Since we are applying cuts to the mixed QG(MC) sample, before we compare it to the G(MC) and Q(MC), we need to see if the cuts themselves distort the studied ratio. In order to see this, from each G(MC) and Q(MC) we select a subset of jets passing the selection criteria either for quark-- or gluon--like jets. From G(MC) we select a subset of jets passing gluon--like criteria (G(sel) $\in$ G(MC)) and from Q(MC) we select a subset of jets passing quark--like criteria (Q(sel) $\in$ Q(MC)). The comparison of these subsets with the inclusive G(MC) and Q(MC) is shown in Fig.\ \ref{fig:fig4}, in the middle. We see that, even though we did select a very narrow fragmentation--class of jets, the ratios are comparable with the MC samples. This already suggests that in Pythia it is rather the origin than the fragmentation property of jet itself which determines the identified particle spectra inside a jet.

Finally, we proceed with the comparison of the Q(sel) and G(sel) samples with the Q(MC) and G(MC). The comparison is shown in the very right panel of Fig. \ref{fig:fig4}. We observe that the gluon--like selection works fine, however, the selected quark--like jets have the ratio closer to G(MC) than to Q(MC). The reason lies in the high contamination of the Q(sel) with the real gluons. We are at energies where the processes producing two gluons in the final state of a hard-scattering have highest probability to occur \cite{Pochybova:2011zzb} and this gives them a high probability to pass our quark selection criteria, thus contaminating the quark--like jets. This supports our statement from previous paragraph, where we say that in the particular MC model which we use it is rather the origin of the jet than its fragmentation that determines the final particle spectra.

\section{Discussion and conclusions}

In this work we have motivated a novel study of particle production in jets, based on their fragmentation, in order to complement other studies focusing on the origin of jets. We have concluded that despite the different fragmentation of jets originating from quarks or gluons, it has no influence on the final particle spectra inside these jets -- within the used MC model. To see whether this statement holds, we suggest to proceed with similar study on experimental data, investigating the dependence of the $p/\pi$ ration inside jets on their fragmentation.

\section*{acknowledgements}
This work was funded by the OTKA Grant NK77816 and NK106119.

I would like to personally thank Tam\'as S\'andor B\'ir\'o and P\'eter L\'evai for valuable consultations.

\end{document}